\documentclass{revtex4}
\usepackage{psfig}
\usepackage{subfigure}
\usepackage{graphicx}
\usepackage{doublespace}
\usepackage{amsmath}
\usepackage{amsfonts}

\begin{document}

\author{Francesco Sorrentino${}^{\ddagger}$, Pietro DeLellis${}^{*}$ }
\affiliation{${}^\ddagger$ Universit{\`a} degli Studi di Napoli Parthenope, 80143 Napoli, Italy \\${}^*$ Universit{\`a} degli Studi di Napoli Federico II, 80127 Napoli, Italy}

\begin{abstract}
{This paper deals with adaptive synchronization of chaos in the presence of time-varying communication-delays. We consider two bidirectionally coupled systems that seek to synchronize through a signal that each system sends to the other one and is transmitted with an unknown time-varying delay. We show that an appropriate adaptive strategy can be devised that is successful in dynamically identifying the time-varying delay and in synchronizing the two systems. The performance of our strategy with respect to the choice of the initial conditions and the presence of noise in the communication channels is tested by using numerical simulations. {Another advantage of our approach is that in addition to estimating the communication-delay, the adaptive strategy could be used to simultaneously identify other parameters, such as e.g., the unknown time-varying amplitude of the received signal.}}
\end{abstract}

\title{Estimation of communication-delays through adaptive synchronization of chaos}
\maketitle


\section{Introduction}

{Over the last twenty years, synchronization of chaos has attracted much attention from the scientific community \cite{FujiYama83,Replace,Pe:Ca,Ding:Ott,SUCNS,Report2}. An interesting result is the observation that two identical chaotic systems, starting from different initial conditions can be synchronized on a stable chaotic time-evolution \cite{He:Va,Kap,CSF1,CSF2,CSF4}. Even if the systems are slightly non-identical, they may nonetheless converge onto an approximately synchronized chaotic time-evolution \cite{Ash1,Ash2,Gh:Ch,CSF3,CSF5,restr_bubbl,Su:Bo:Ni,So:Po}. {Stochastic synchronization for two coupled systems or for an arbitrary network in the presence of time delays has been studied in  \cite{Hu:Ko:Sz10,Hu:Ko:Sz11}.} Applications of synchronization of chaos include but are not limited to secure communication \cite{Cu:Op,Carr2,Argyris,Feki,CSF6}, system identification \cite{CSF7,Abarbanel, Abarbanel2,Abarbanel4,IDTOUT}, data assimilation \cite{So:Ott:Day,Duane}, sensors \cite{SOTT2}, 
information encoding and transmission \cite{Ha:Gr:Ott,Dr:He:An:Ott}, and multiplexing \cite{Ts:Su}. In this paper, we show that two chaotic systems can be synchronized in the presence of  a time-varying unknown communication-delay through an appropriate adaptive strategy; moreover, by using this adaptive strategy, each system is able to independently formulate a dynamical estimate of the unknown communication-delay.} 

It was proposed in \cite{OGY} that the property of a system of being chaotic can be conveniently exploited in control applications. In this paper, we consider an identification problem where an unknown time-varying communication-delay between two bidirectionally coupled systems 
is to be estimated. We show that choosing the dynamics of the individual systems to be chaotic is convenient in terms of the identification strategy. Moreover, by using our strategy we are able to isochronally synchronize the two systems.

{As a reference application, we consider the problem of identifying  a time-varying communication-delay between two autonomous moving platforms, though we aim at presenting a general methodology rather than a specific application.
Moreover, our paper provides new insight into the phenomenon of synchronization of chaos, as we are the first ones to formulate an adaptive strategy that successfully addresses synchronization of chaos and identification of time-varying communication-delays altogether.}

In Sec.\ II we present the formulation of our problem and as a reference application, we introduce the problem of identifying  a time-varying communication-delay between two autonomous moving platforms. In Sec.\ III we formulate an adaptive strategy to identify unknown communication-delays in the case of non-autonomous systems, which is validated by using numerical simulations. In Sec.\ IV we show that a \emph{decentralized agreement protocol} can be used together with our adaptive strategy to identify the coupling-delays  for the case of autonomous systems. The effects of the choice of the initial conditions are investigated in Sec.\ V. In Sec.\ VI we present a modified adaptive strategy that performs better in the presence of noise in the communication channels.  Finally, the conclusions are given in Sec.\ VII.

\section{Formulation}

{A typical scheme for synchronization of two chaotic systems in the presence of communication-delays is the following \cite{DelayChen,DelayChen2,DelayZhou,Ki:En:Re:Zi:Ka}},
\begin{equation}\label{general}
\dot{x}_i(t)= F(x_i(t),t)+\Gamma_i[H(x_j(t-\tau_i))-H(x_i(t-\tau_i))],
\end{equation}
$i=\{1,2\}$, $j=(3-i)$. $F : \mathbb{R}^n \times \mathbb{R}^+ \rightarrow \mathbb{R}^n$ describes the dynamics of each uncoupled chaotic system, $H : \mathbb{R}^n \rightarrow \mathbb{R}$ is a scalar output function, $\Gamma_i$ is an $n$-dimensional constant coupling vector, the communication-delay $\tau_i$ is the time that it takes for the signal broadcast by node $j$ to be received by node $i$. The delays $\tau_i$  in Eq. (\ref{general}) are assumed constant and known.

{Note that Eqs. (\ref{general}) are a system of delay differential equations. Hence the solution $x_i(t)$, $i=\{1,2\}$, is determined by knowledge of the initial conditions for $x_i$ over the time-interval $[-\tau_{max},0]$, where $\tau_{max}=\max(\tau_1,\tau_2)$.
}

In the particular case in which $x_1=x_2=x_s$, the terms in the square brackets of (\ref{base}) vanish, and the two systems evolve on the synchronous solution,
\begin{equation}
\dot{x}_s(t)=F(x_s(t),t).\label{s}
\end{equation}
For a given choice of the functions $F$ and $H$, stability of the synchronized solution depends on the parameters $\Gamma_1$, $\Gamma_2$, $\tau_1$, and $\tau_2$.
In particular, a condition for stability is that the communication-delays $\tau_i$ be smaller than the characteristic timescale $T_x$ of  an uncoupled system (\ref{s}) \cite{Ki:En:Re:Zi:Ka},
\begin{equation}\label{TS1}
\tau_i \leq T_x,
\end{equation}
$i=\{1,2\}$. 

If we assume that the communication-delay from $i$ to $j$ is the same as that from $j$ to $i$, Eq. (\ref{general}) reduces to
\begin{equation}\label{base}
\dot{x}_i(t)= F(x_i(t),t)+\Gamma_i[H(x_j(t-\tau))-H(x_i(t-\tau))],
\end{equation}
$i=\{1,2\}$, $j=(3-i)$, $\tau_1=\tau_2=\tau$. The assumption that the communication-delays are the same in both directions holds true for example for those applications that use line of sight communication.

In this paper, we choose the function $F$ in (\ref{s}) to generate synchronous (uncoupled) chaotic dynamics. Chaotic signals have been successfully employed in cryptography and in secure communication \cite{Cu:Op,Ann:96,Ban:09}. Adaptive strategies based on synchronization of chaos have also been proposed  to dynamically identify the parameters of unknown systems \cite{Abarbanel, Abarbanel2,Abarbanel4,IDTOUT}.  In such applications, it is possible to exploit the specific properties of chaotic signals. In fact, since a chaotic time trace never repeats in time, it provides an infinite amount of information that can be used by the identification strategy (see e.g., \cite{IDTOUT,IDDD}). {Moreover, adaptive strategies have been devised to synchronize self-sustained chaotic systems \cite{Ka:Bo:Se:Ku}.}

In this paper, we propose an adaptive strategy {that exploits the synchronizability property of chaotic systems} to estimate a time-varying unknown communication-delay. Our goal is twofold:

  i) synchronizing the two systems;

  ii) dynamically and independently estimating at each system the communication-delay.

To our knowledge, the use of adaptive strategies to simultaneously achieve synchronization and estimate unknown communication-delays has not been previously addressed in the literature. A strategy based on synchronization of chaos to identify the unknown strengths of the couplings between two or more coupled systems has been proposed in \cite{SOTT} and implemented in \cite{EXP,MurphyEtAl,EXP2}. Synchronization of chaos has also been used to identify and predict the dynamics of unknown real systems \cite{Abarbanel,Abarbanel2,Abarbanel4,IDTOUT}. Adaptive strategies have been proposed to enhance synchronization of coupled dynamical systems by acting on the coupling strengths \cite{Zh:Ku06,Delellis3,Zh:Zh} and the topology of interconnections \cite{Delellis4}.
Unknown parameters have been adaptively estimated in the presence of time-delays in \cite{Yu:Parlitz,Yu:Cao07,Yu:Cao08,Identif1,Identif_S2N}. A technique to estimate the internal delay of an unknown system has been proposed in \cite{IDDD}. The same problem has been studied in \cite{Delay_ref1} for the case of linear systems. 
In a recent paper \cite{Ja:Ha:Ta}, synchronization of chaos in the presence of an unknown communication-delay is studied between two unidirectionally coupled systems, whose synchronization is guaranteed by a connection to an external master system. In this case, the problem of synchronizing the two systems is separate from that of identifying the delay. In this paper, we consider two bidirectionally coupled systems, we do not assume connection to an external master system, 
and we formulate an adaptive strategy that can be used to simultaneously synchronize the two systems and identify the delay.


{In what follows we present the adaptive synchronization strategy in terms of a reference application. In particular, we consider the problem of identifying  a time-varying communication-delay between two autonomous moving platforms. We assume that two identical chaotic oscillators are installed at each platform and that these chaotic systems seek to synchronize via a signal broadcast from one platform to the other (and viceversa). The signal received at each platform is transmitted with a time-varying delay, which depends on the relative distance between the two platforms. Information on the communication-delay can be used, e.g., to estimate the relative distance.
}

Each platform is characterized by a pair of state variables $\{p_i(t),x_i(t)\}$, $i=\{1,2\}$, where $x_i(t) \in \mathbb{R}^n$ is the state of an oscillator installed at platform $i$, and $p_i(t) \in \mathbb{R}^m$ is the position of the platform $i$. We assume that the two platforms seek to achieve mutual synchronization of the $x$'s dynamics (the dynamics of the oscillators), while independently moving along the trajectories $p_i(t)$. In order to achieve synchronization, each platform broadcasts a signal which encodes information on the evolution of the $x$'s dynamics. The signals are received  with a non-negligible communication-delay, which depends on the relative motion of each platform with respect to each other.

We consider that each uncoupled oscillator is described by $\dot{x}_i(t)=F(x_i(t),t)$. Moreover, we assume that $p_i(t)$ typically evolves on a timescale $T_p$ which is much longer than the timescale $T_x$ on which an uncoupled system evolves, that is,
\begin{equation}\label{TS2}
T_p \gg T_x.
\end{equation}
This may be achieved, for example, by choosing the individual oscillators to have very fast dynamics, i.e., much faster than the dynamics of $p_i(t)$, depending also on the specific application of interest.  System $i=\{1,2\}$ broadcasts a signal $H(x_i(t))$, which is received by $j=(3-i)$ with a delay $\tau_i(t)$, which is a function of both $p_i(t)$ and $p_j(t)$. Under our assumption that $T_p \gg T_x$, we can approximate $ \tau_1(t), \tau_2(t)$ with the same delay, say $\tau(t)$, with $\tau(t)$ being a function of the distance $\| p_i(t)-p_j(t) \|$.

When the platforms are coupled, the equations for the oscillators become,
\begin{subequations}\label{main}
\begin{align}
\dot{x}_i(t)= & F(x_i(t),t)+\Gamma_i[r_i(t)-H(x_i(t-\tau'_i(t)))],  \\
r_i(t)= & H(x_j(t-\tau(t))),
\end{align}
\end{subequations}
$i=\{1,2\}$, $j=(3-i)$. Here, $F,H$, and $\Gamma_i$ are the same as in Eq. (\ref{base}). The received signal at node $i$, $r_i(t)$, propagates from node $j$ to node $i$ with a communication-delay $\tau(t)$. Note that in Eqs. (\ref{main}a), we subtract from  the {received signal} $H(x_j(t-\tau(t)))$ the {internal signal} $H(x_i(t-\tau'_i(t)))$, $i=\{1,2\}$ , where $\tau'_i(t)$ is an estimate of $\tau(t)$ at node $i$.  In the particular case in which $x_1=x_2=x_s$ and $\tau'_1=\tau'_2=\tau$, the terms in the square brackets of (\ref{main}a) vanish, and the two systems evolve on the synchronous solution (\ref{s}). {Note that in this case, integration of the system of equations (\ref{main}) requires knowledge at each time $t$ of the state variables $x_i$ over the time-interval $[t-\tau_{max}(t),t]$, where $\tau_{max}(t)=\max(\tau(t),\tau'_1(t),\tau'_2(t))$.  }

{A fundamental issue that typically arises when considering communication between two or more  autonomous moving platforms is that of synchronization of the time-clocks at each platform \cite{AVIO1,AVIO2}. For example, the problem of synchronization of non-autonomous chaotic systems has been studied in \cite{Carr1}. In Sec.\ III   we will assume the presence of two perfectly synchronized internal clocks at the two platforms. This is reflected by the choice of non-autonomous chaotic oscillators at the two platforms, $\dot{x}_i(t)= F(x_i(t),t)$, for which the dynamics at the two oscillators explicitly depends on the same variable $t$. As we will see, this requirement is important for the successfulness of the adaptive strategy, i.e., for simultaneously synchronizing the two systems and correctly estimating the unknown communication-delays. However, in Sec.\ IV we  show that a \emph{decentralized agreement protocol} can be used to estimate the delay and to synchronize the two oscillators for the case that they are autonomous systems.}

\subsection{Uniqueness of the solution for both the cases of autonomous and non-autonomous oscillators}

We now consider that the uncoupled dynamics at each platform is autonomous, i.e., we replace $F(x_i(t),t) \rightarrow F(x_i(t))$ in Eq. (\ref{main}a). We show that when such a modification to our scheme is done, we become unable to univocally determine the unknown delay $\tau$, by requiring that \emph{the two systems synchronize}, i.e., that   $H(x_j(t-\tau))=H(x_i(t-\tau'_i))$ in Eqs. (\ref{main}), $i=\{1,2\}$, $j=(3-i)$.

A chaotic time trace never repeats in time. Since $H(x_i(t))$ in (\ref{main}) is chaotic, $i=\{1,2\}$, the only possibility for the received signal $H(x_j(t))$ to cancel out with the internal signal $H(x_i(t))$ over time is that the two chaotic time traces are the same one, possibly translated by a fixed time lag equal to $\Delta$, i.e.,
\begin{equation}
x_1(t)=x_2(t-\Delta). \label{Delta}
\end{equation}
{An example of $\Delta$-lag synchronization between two chaotic time series (Eq. (\ref{Delta})) is shown in Fig.\ 1. Our goal in this paper will be setting $\Delta$ to $0$. Fig.\ 2 shows an example of lag synchronization between two identical $P$-periodic time series. As can be seen, for this case, lag-synchronization is possible for value of the lags equal to $\Delta + N P$, with $N$ being any integer number. Thus using chaotic dynamics at the individual nodes restricts the possibility of lag-synchronization to only one lag $\Delta$. 

By setting the coupling terms in the square brackets of Eqs. (\ref{main}) to zero, and using (\ref{Delta}), we obtain,
\begin{subequations}\label{tre}
\begin{align}
x_2(t-\tau)=x_2(t-\tau'_1-\Delta), \\
x_2(t-\tau-\Delta)=x_2(t-\tau'_2).
\end{align}
\end{subequations}
Again, by observing that $x_2(t)$ is chaotic, Eqs. (\ref{tre}) reduce to,
\begin{subequations}\label{relt}
\begin{align}
\tau=\tau'_1+\Delta, \\
\tau+\Delta=\tau'_2.
\end{align}
\end{subequations}

In Eqs. (\ref{relt}), $\tau$ is a  time-varying external parameter. Then we see from (\ref{relt}) that for each value of $\tau$, the two systems may show $\Delta$-lag synchronization if the following conditions are satisfied,
\begin{subequations}\label{relt2}
\begin{align}
\tau=\tau'_1+\Delta, \\
2 \Delta=\tau'_2-\tau'_1. \label{gen}
\end{align}
\end{subequations}
A case of interest is that equations (\ref{relt2}) are satisfied for,
\begin{equation}
\tau'_2=\tau'_1=\tau, \label{goal}
\end{equation}
corresponding to $\Delta=0$. Condition (\ref{goal}) corresponds to a particular (desirable) solution for (\ref{relt2}), for which both the estimates $\tau'_1$ and $\tau'_2$ converge on the \emph{true} value of $\tau$, with the two systems synchronizing with lag equal zero.

\begin{figure}[t]
\centering
\includegraphics{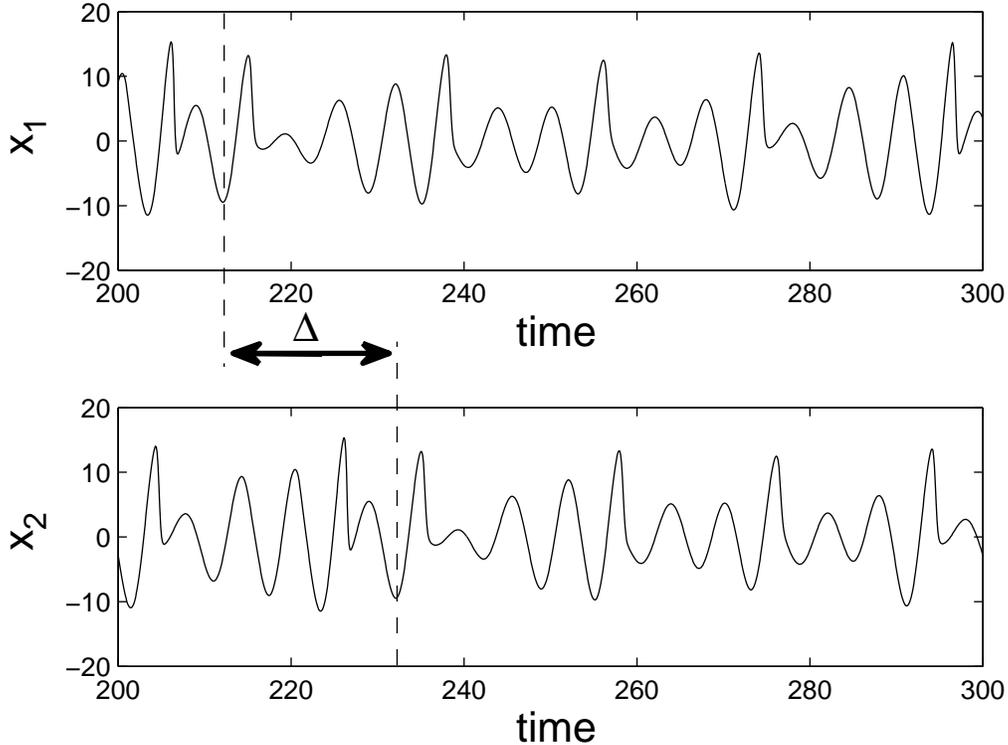}
\caption{\small The figure shows an example of $\Delta$-lag synchronization between two chaotic time series generated by integrating Eqs. (\ref{ROSS}). Our goal in this paper we will be setting $\Delta$ to $0$. }
\end{figure}
\begin{figure}[h!]
\centering
\includegraphics{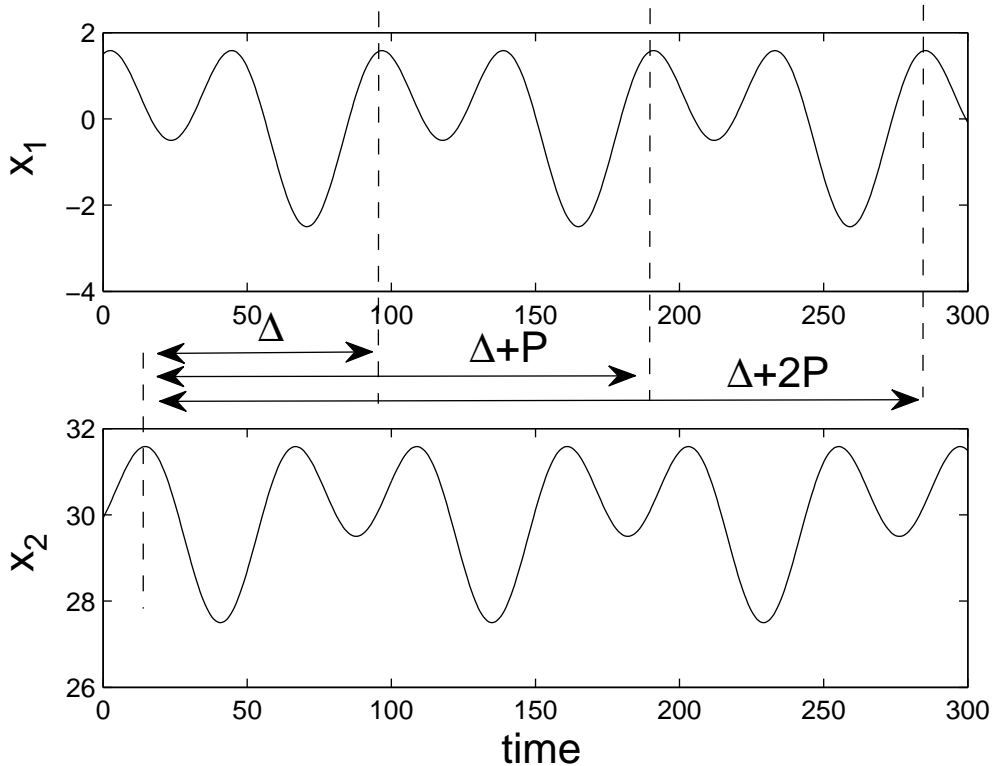}
\caption{\small The figure shows an example of lag synchronization between two $P$-periodic time series. As can be seen from the figure, lag synchronization can be observed for lags  equal to  $\Delta + N P$, with $N$ being any integer number. }
\end{figure}

 In what follows, we will propose an adaptive strategy to evolve the parameters $\tau'_i$ in such a way to minimize the coupling terms in the square brackets of Eqs. (\ref{main}). However, we should note that by setting to zero the coupling terms in the square brackets of Eqs. (\ref{main}), a general solution of the type (\ref{relt2}) will arise, and not necessarily of type (\ref{goal}).

Thus to avoid ambiguities in resolving the coupling
delays, hereafter we consider two alternative possible remedies, one of which applies to nonautonomous systems and the other one applies to autonomous systems. As a first solution, we assume that the dynamics of the uncoupled $x$'s in Eqs. (\ref{main}) are non-autonomous, i.e., they explicitly depend on the time variable $t$. Then, if such dependence on $t$ is appropriately chosen, under the assumption of chaotic dynamics for the $x$'s, we can rule out the possibility that Eq. (\ref{Delta}) is satisfied for $\Delta \neq 0$. To better explain our point, let us recur to an example. Consider a simple periodic dependence on time of the type, $\dot{x}_i(t)=F(x_i(t),\sin(2 \pi \lambda t))$; then the only way Eqs. (\ref{Delta}) can be satisfied is for $\Delta$ being an integer multiple of ${\lambda}^{-1}$, i.e., $\Delta=\{0,{\lambda}^{-1},2 {\lambda}^{-1},...\}$. However, not all of these solutions will be stable, with stability being typically bounded by a maximum allowed lag $\Delta$, say $\Delta_{max}$ (which depends on the autocorrelation time for an uncoupled system, see \cite{Ki:En:Re:Zi:Ka}). Therefore, by choosing ${\lambda}^{-1}>\Delta_{max}$, we can ensure that the only possible stable solution of type (\ref{Delta}) is for $\Delta=0$. {The advantage of this approach is that  it uses the received signal (\ref{main}b) as the only available information at each platform  and from that it is able to reconstruct the communication-delay with which the signal propagates from the receiver to the sender. The disadvantage is that it requires the presence of two perfectly synchronized internal clocks at the two platforms.}

{A second possible solution ensuring $\Delta=0$ in Eqs. (\ref{relt2}) is to set $\tau'_2=\tau'_1$. In fact, with this condition, Eqs.  (\ref{relt2}) yields Eq. (\ref{goal}). 
However, we note that combining information available at the two different platforms to satisfy the condition that $\tau'_2$ be equal to $\tau'_1$ may be unpractical. In Sec.\ IV, we introduce a \emph{decentralized agreement protocol} with this specific purpose.  The advantage of this alternative strategy is that it allows to work with autonomous dynamical systems at the two platforms,  rather than non-autonomous systems. The disadvantage is that it requires more information exchanged between the two platforms. 
}



\section{Adaptive Strategy}

Our goal is independently estimating at each platform $i$ the unknown communication-delay $\tau(t)$ from sole knowledge of the received signal $r_i(t)$, $i=1,2$. To this aim, we present an adaptive strategy to evolve $\tau'_i(t)$, the estimate at platform $i$,  to match the unknown communication-delay $\tau(t)$.
At each node $i$, we consider a \emph{potential},
\begin{equation}\label{P}
\Psi_i(t)=[H(x_j(t-\tau(t)))-H(x_i(t-\tau'_i(t)))]^2.
\end{equation}
From the considerations in Sec.\ IIA, we see that $\Psi_i=0$ if (i) $x_j=x_i$ and (ii) $\tau'_i=\tau$, $i=1,2$. Therefore, we seek to evolve our estimates $\tau'_i$ to minimize $\Psi_i$, through the following gradient descent relations,
\begin{equation} \label{adt}
\dot{\tau'}_i(t)=-\alpha_i \frac{\partial \Psi_i}{\partial \tau'_i}=-2 \alpha_i [H(x_j(t-\tau(t)))-H(x_i(t-\tau'_i(t)))] DH(x_i(t-\tau'_i(t))) \dot{x}_i(t-\tau'_i(t)),
\end{equation}
$\alpha_i>0$, $DH$ represents the derivative of the function $H$ with respect to its argument. Note that (\ref{adt}) involves knowledge of the received signal $r_i(t)=H(x_j(t-\tau(t)))$, the state $x_i$ at the past time $(t-\tau'_i)$,  and its derivative $\dot{x}_i$  calculated at time $(t-\tau'_i)$.

{We note here that our proposed strategy, Eqs. (\ref{main},\ref{adt}), involves the solution of  non-autonomous delay differential equations with state-dependent delays. 
In this paper, the stability of our strategy is tested via numerical simulations and provides interesting insights on the viability of our proposed approach. Namely, we show its effectiveness for different systems, given that the initial guess on the time delay is sufficiently close to the actual initial delay. Furthermore, the sensitivity to the choice of the initial conditions and to the presence of noise in the communication channels is investigated in Secs.\ V and VI.}

\subsection{Numerical Experiments}

In what follows, we consider a numerical experiment, for which we choose the dynamics of each uncoupled system $\dot{x}_i=F(x_i(t),t)$ to be described by the forced Van der Pol oscillator.
We restrict the parameters of the coupled equations to be such that with no adaption being performed and 
$\tau'_1(t)=\tau'_2(t)=\tau(t)$,
the synchronous solution (\ref{s}) is stable. This also corresponds to setting a maximum value on the communication-delay $\tau(t)$, as stability requires that $\tau$ does not exceed the autocorrelation time of an uncoupled system (for more details, see \cite{Ki:En:Re:Zi:Ka}).

\begin{figure}[t]
\centering
\includegraphics{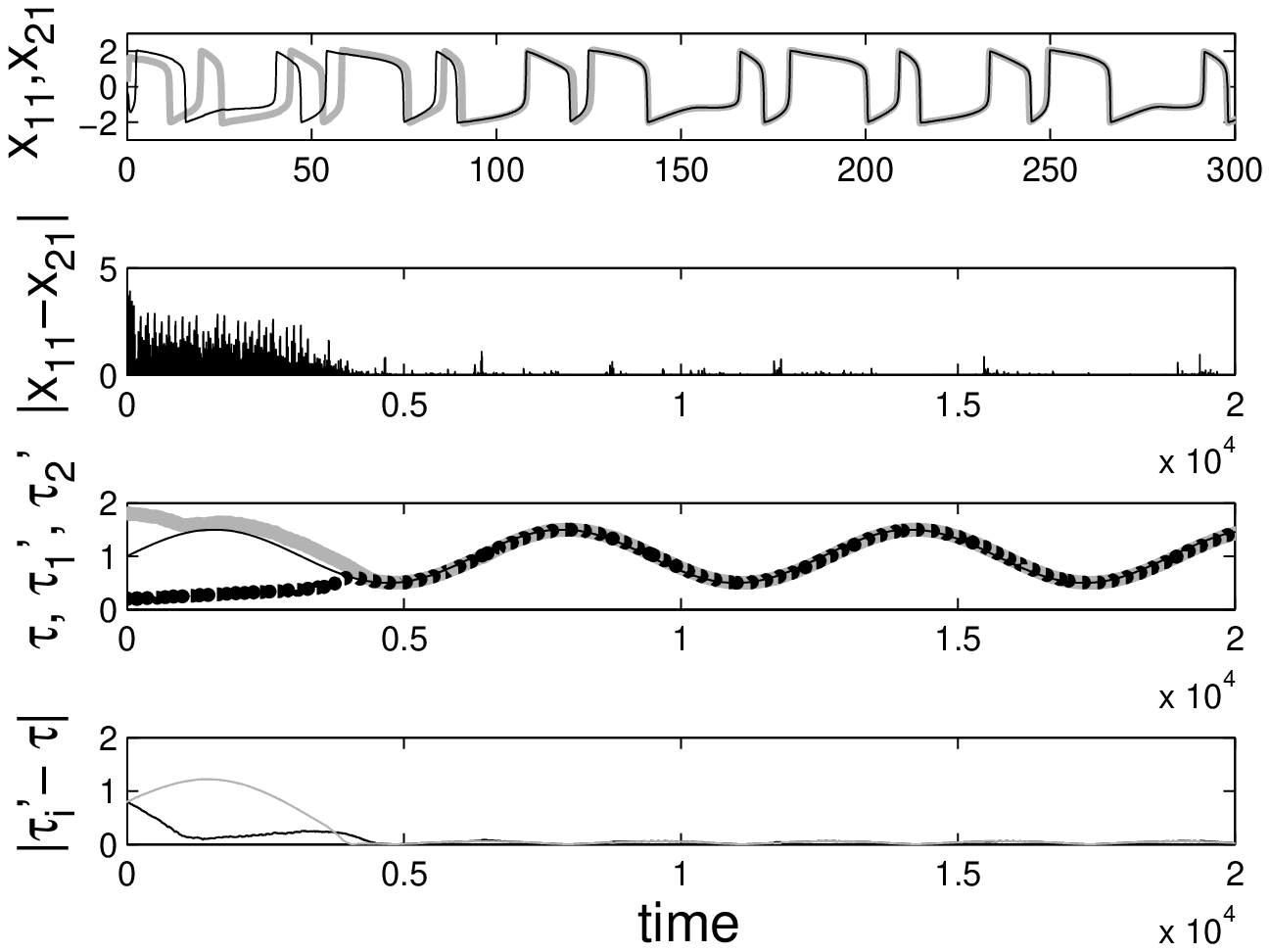}
\caption{\small We integrate the set of equations (\ref{main},\ref{adt},\ref{VDP}) with $\tau$ evolving according to (\ref{taut}), $T=1$, $\zeta=0.5$, $\omega=10^{-3}$. We set $\alpha_1=\alpha_2=10^{-4}/2$,
$\gamma_1=\gamma_2=0.15$. From top to bottom: time evolution of $x_{11}(t)$ and $x_{21}(t)$; synchronization error $|x_{11}(t)-x_{21}(t)|$; time evolution of $\tau(t)$ (thin black curve) $\tau'_1(t)$ (thick dashed grey curve), and $\tau'_2(t)$ (thick dotted black curve); time evolution of the delay estimation errors, $|\tau(t)-\tau'_1(t)|,|\tau(t)-\tau'_2(t)|$. We specify the initial conditions for (\ref{main}) to be constant over the time-interval $[-\bar{\tau}_{max},0]$, where $\bar{\tau}_{max}=\max_t \tau_{max}(t)$, and equal to randomly chosen points on the forced Van der Pol chaotic attractor. Moreover, we set $\tau'_1(0)=1.8$, $\tau'_2(0)=0.20$.  }
\end{figure}

We set $x_i(t)=[x_{i1}(t),x_{i2}(t)]$,
\begin{equation}
F(x_{i1}(t),x_{i2}(t),t)= \left[
  \begin{array}{cc}
    x_{i2}(t) \\
     -x_{i1}(t)+8.53 x_{i2}(t) [1-x_{i1}(t)^2]+f(t)  \\
  \end{array}
\right], \label{VDP}
\end{equation}
where,
\begin{equation} \label{f}
f(t)= a \cos(2 \pi \mu t),
\end{equation}
$a=1.2$, $\mu=0.1$, and we set $H(x_i(t))=x_{i2}(t)$, $\Gamma_i=[0,\gamma_i]$. 

We integrate the set of equations (\ref{main},\ref{adt}), with
\begin{equation}
\tau(t)=T+\zeta \sin(\omega t), \label{taut}
\end{equation}
$\omega=10^{-3}$ and we set $T=1$, $\zeta=0.5$. {We set the initial conditions for (\ref{main}) to be constant over the time-interval $[-\bar{\tau}_{max},0]$, where $\bar{\tau}_{max}=\max_t \tau_{max}(t)$ and equal to randomly chosen points on the forced Van der Pol chaotic attractor. We set $\tau'_1(0)=1.8$, $\tau'_2(0)=0.20$ (note that integration of Eqs. (\ref{main},\ref{adt}) only requires knowledge of the delays $\tau'_i$ at the current time $t$).}

The results of our computations are shown in Fig.\ 3. 
As can be seen, the adaptive strategy is successful in synchronizing the oscillators' states $x_1$ and $x_2$ on a zero-lag synchronous solution, with $\tau'_1(t),\tau'_2(t)$ converging on the true evolution $\tau(t)$.

\section{A decentralized agreement protocol for autonomous systems}

{In this section we focus on the case that the two coupled oscillators are autonomous systems, i.e., we replace $\dot{x}_i(t)=F(x_i(t),t)$ in Eqs. (\ref{main}) by $\dot{x}_i(t)=F(x_i(t))$. For this case, the existence of only one solution of type (\ref{goal}) can be ensured by satisfaction of the extra condition that $\tau'_1=\tau'_2$   (see the discussion in Sec.\ IIA). In order to satisfy this additional requirement, we introduce a  decentralized agreement protocol between the two platforms that consists in each oscillator $i$ communicating to oscillator $j=(3-i)$ its estimate $\tau'_i$ of the unknown delay. 
With this modification, each platform $j$ receives at each time two signals, that is, the signal $r_j=H(x_i(t-\tau(t)))$ and the delay estimate $\tau'_i(t-\tau(t))$.
With this extra piece of information available, we reformulate the potential in (\ref{P}) as follows,
\begin{equation}\label{PH}
\hat{\Psi}_i(t)=\Psi_i(t)+\kappa [\tau'_j(t-\tau(t))-\tau'_i(t-\tau'_i(t))]^2,
\end{equation}
$i=\{1,2\}$, $j=(3-i)$, where $\kappa>0$ is an appropriate scalar. Note that in defining (\ref{PH}), we have 
taken into account the transmission-time $\tau$ needed for the estimate $\tau'_j$ to propagate from platform $j$ to platform $i$. We note that, as we see from the relations (\ref{TS1},\ref{TS2}), $\tau$ evolves on a timescale that is much longer than $\tau$ itself, thus we do not expect it to vary much over the propagation-time of the signal.
 We observe from (\ref{PH}) that the potential $\hat{\Psi}_i \geq 0$ by definition. Then, in order for the potential (\ref{PH}) to be zero,  it is simultaneously needed that  $\Psi_i$ equals zero and $\tau'_j(t-\tau)$ equals $\tau'_i(t-\tau'_i)$.
Thus we introduce the following gradient descent relation for evolving the estimates $\tau'_i$,
\begin{equation} \label{adtH}
\dot{\tau'}_i=-\alpha_i \frac{\partial \hat{\Psi}_i}{\partial \tau'_i}=-2 \alpha_i  \{ [H(x_j(t-\tau(t)))-H(x_i(t-\tau'_i(t)))] DH(x_i(t-\tau'_i(t))) \dot{x}_i(t-\tau'_i(t)) -\kappa [\tau'_j(t-\tau(t))-\tau'_i(t-\tau'_i(t))] \},
\end{equation}
$\alpha_i>0$. 
Note that (\ref{adtH}) involves knowledge of the two received signals $H(x_j(t-\tau(t)))$ and $\tau'_j(t-\tau(t))$, along with knowledge of the state $x_i$, of its derivative $\dot{x}_i$, and of the estimate $\tau'_i$ calculated at the past time $(t-\tau'_i)$.
}
The results of our computations are shown in Fig.\ 4. We choose the dynamics of each individual system to be described by the R\"ossler autonomous equation,  $x_i(t)=[x_{i1}(t),x_{i2}(t),x_{i3}(t)]$,
\begin{equation}
F(x_{i1}(t),x_{i2}(t),x_{i3}(t),t)= \left[
  \begin{array}{ccc}
    & -x_{i2}(t)-x_{i3}(t) \\
    & x_{i1}(t)+0.2 x_{i2}(t) \\
    & {x_{i3}(t)}[x_{i1}(t)-7]+0.2 \\%
  \end{array}
\right]. \label{ROSS}
\end{equation}
 We integrate the set of equations (\ref{main},\ref{ROSS},\ref{adtH}) with $H(x_i(t))=x_{i1}(t)$, $\Gamma_i=[\gamma_i,0,0]$. The unknown delay $\tau$ evolves according to Eq. (\ref{taut}) with  $T=0.3$, $\zeta=0.15$, $\omega=10^{-3}$. We set $\alpha_1=\alpha_2=10^{-4}/2$, $\gamma_1=0.6$, $\gamma_2=0.5$, and $\kappa=1$. We select the initial conditions for  $x_{i1},x_{i2},x_{i3}$  to be constant over the time-interval $[-\bar{\tau}_{max},0]$, where $\bar{\tau}_{max}=\max_t \tau_{max}(t)$, and equal to randomly chosen points on the R\"ossler chaotic attractor, $\tau'_1(0)=0.65$, $\tau'_2(0)=0.10$. As can be seen, the adaptive strategy along with the decentralized agreement protocol is successful in synchronizing the oscillators' states $x_1$ and $x_2$ on a zero-lag synchronous solution, with $\tau'_1(t)$ and $\tau'_2(t)$ converging on the true evolution  $\tau(t)$.
\begin{figure}[t]
\centering
\includegraphics{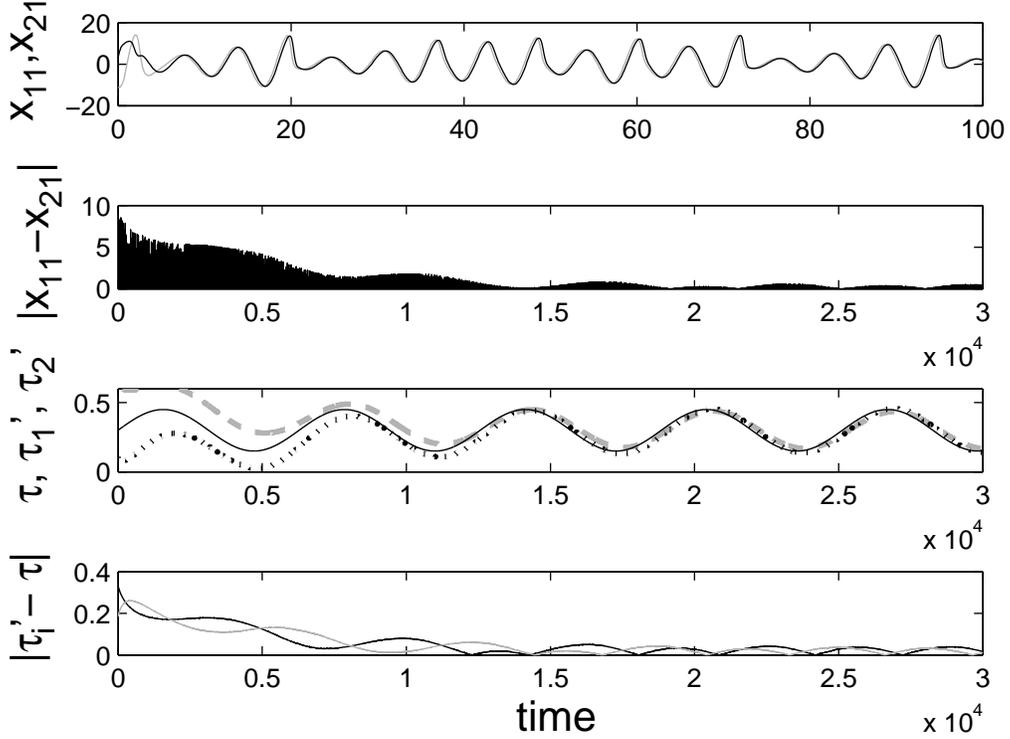}
\caption{\small We integrate the set of equations (\ref{main},\ref{adtH},\ref{ROSS}) with $\tau$ evolving according to (\ref{taut}), $T=0.3$, $\zeta=0.15$, $\omega=10^{-3}$. We set  $\alpha_1=\alpha_2=10^{-4}/2$,
$\gamma_1=0.6$, $\gamma_2=0.5$, $\kappa=1$. From top to bottom: time evolution of $x_{11}(t)$ and $x_{21}(t)$; synchronization error $|x_{11}(t)-x_{21}(t)|$; time evolution of $\tau(t)$ (thin black curve) $\tau'_1(t)$ (thick dashed grey curve), and $\tau'_2(t)$ (thick dotted black curve); time evolution of the delay estimation errors, $|\tau(t)-\tau'_1(t)|,|\tau(t)-\tau'_2(t)|$. We select the initial conditions for $x_{i1},x_{i2},x_{i3}$ to be constant over the time-interval $[-\bar{\tau}_{max},0]$, where $\bar{\tau}_{max}=\max_t \tau_{max}(t)$, and equal to randomly chosen points on the R\"ossler chaotic attractor, $\tau'_1(0)=0.65$, $\tau'_2(0)=0.10$.  }
\end{figure}

{Furthermore, we have tested the performance of the adaptive strategy described in this section with respect to different sets of parameters. We have run numerical simulations in which we integrate the set of equations (\ref{main},\ref{adtH},\ref{ROSS}) with $\kappa=1$ and $\tau$ evolving according to (\ref{taut}), $T=0.4$, $\zeta=0.2$, $\omega=10^{-3}$.  In order to assess the effectiveness of the adaptive strategy we monitor the estimation error,}
\begin{equation}
E_i(t)=|\tau(t)-\tau'_i(t)|.
\end{equation}
{In Fig.\ 5 we plot the average estimation error,}
\begin{equation} \label{Ei}
<E_i>=10^{-4} \int_{10^4}^{2 \times 10^4} E_i(t) dt,
\end{equation}
{$i=1$, as we vary the coupling gains $\gamma_1=\gamma_2$ between the two systems  and the adaptation gains $\alpha_1=\alpha_2$. As can be seen, the strategy is quite sensitive to the choice of both the coupling gains and the adaptation gains, which suggests that these parameters should be carefully tuned in order to obtain a desired or improved performance.}

\begin{figure}[t] \label{ult}
\centering
\includegraphics{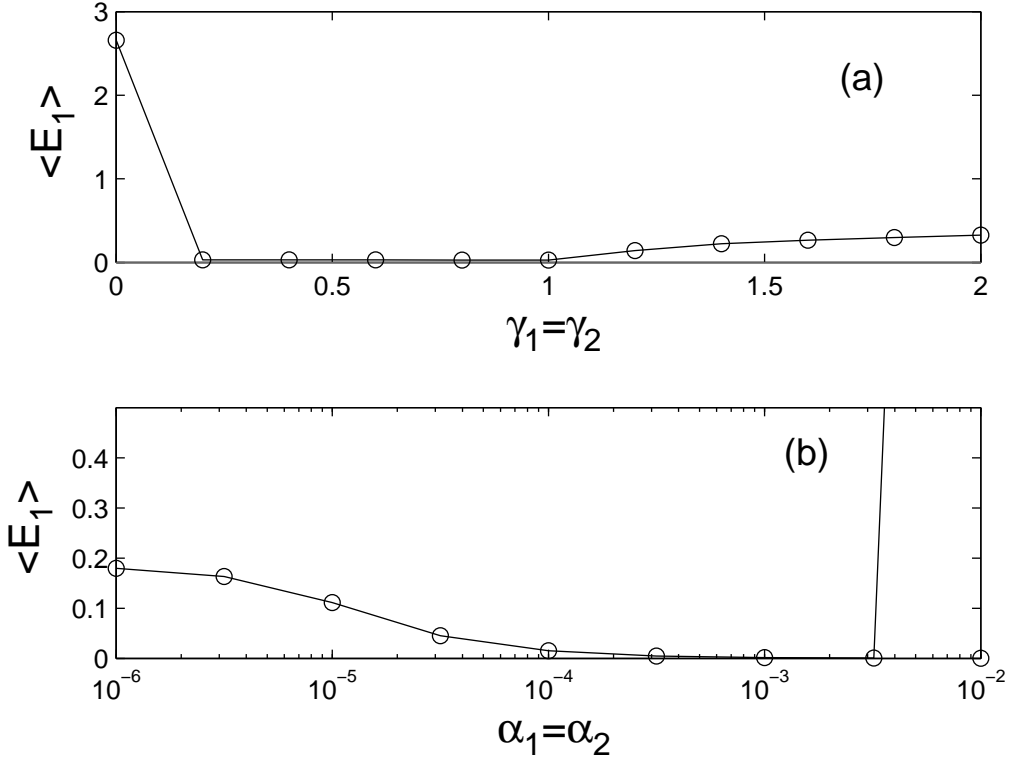}
\caption{\small We integrate the set of equations (\ref{main},\ref{adtH},\ref{ROSS}) with $\tau$ evolving according to (\ref{taut}) with $T=0.4$, $\zeta=0.2$, $\omega=10^{-3}$. We set $k=1$. (a) We plot the average estimation error $<E_1>$ as we vary the coupling gains between the two systems $\gamma_1=\gamma_2$ with $\alpha_1=\alpha_2=10^{-4}/2$.
(b) We plot the average estimation error $<E_1>$ as we vary the adaptation gains $\alpha_1=\alpha_2$ with $\gamma_1=\gamma_2=0.6$. In both plots, each point is an average over many different realizations (i.e., over different choices of the initial conditions for $x_1$ and $x_2$).  }
\end{figure}

{{We observe that the use of the approach proposed in this section is not limited to the reference application of identifying  a time-varying communication-delay between two autonomous moving platforms. Indeed, the version of the adaptive strategy that applies to autonomous systems does not require the presence of synchronized clocks at the two platforms; hence, it could be used to synchronize two autonomous clocks in the presence of clock drifts or skews (see \cite{CLOCK} for a survey on clock synchronization) or to estimate propagation delays in an asynchronous code-division multiple access communication system \cite{CDMA}. As another possible application, our setup could be used as a bistatic radar, where one platforms (acting as a transmitter) communicate a signal to the other one (acting as a receiver) in the presence of a moving target. In this case, knowledge of the time delay for the signal to bounce off the moving target and reach the receiver could be used to extract information on the position of the target.}}


\section{Effects of the Initial Conditions}

 In this section, we investigate the effect of the choice of the initial conditions $\{\tau'_1(0),\tau'_2(0)\}$ on the successfulness of our adaptive strategy. We consider both the forced Van der Pol equation (\ref{VDP}) as well as the forced R\"ossler equation (\ref{ROSS}) to describe the dynamics of an uncoupled system. 
 We choose $\tau'_1(0)=\tau(0)+\Delta_1$, $\tau'_2(0)=\tau(0)+\Delta_2$, where $\Delta_i$ is a measure of how off our initial guesses $\tau'_i$ are with respect to the true value of $\tau(0)$. We select the initial conditions for $x_i$ to be constant over the time-interval $[-\bar{\tau}_{max},0]$, where $\bar{\tau}_{max}=\max_t \tau_{max}(t)$, and equal to randomly chosen points on the R\"ossler chaotic attractor. We run numerical simulations in which equations (\ref{main},\ref{adt},\ref{VDP}) are integrated for a long time, $H(x_i(t))=x_{i2}(t)$, $\Gamma_i=[0,\gamma_i]$, $\alpha_1=\alpha_2=10^{-4}/2$, and
$\gamma_1=\gamma_2=0.15$. At each time $t$, we monitor the estimation error $<E_2>$ defined in Eq. (\ref{Ei})
versus the initial mismatch $\Delta_2$ for different values of $\Delta_1$: circles correspond to $\Delta_1=-0.9$, squares correspond to $\Delta_1=0$, and asterisks correspond to $\Delta_1=0.9$. Each plotted point is averaged over many different realizations, i.e., over different choices of the initial conditions for $x_1$ and $x_2$ (constant over the time-interval $([-\bar{\tau}_{max},0])$ ).
As can be seen, for $0 \leq \Delta_2 \lesssim 1.2$, our strategy is able  to produce good estimates of the unknown time-varying delay $\tau(t)$ within the time-span of our simulations ($2 \times 10^4$). We remark that for $\Delta_1$ in the interval $-0.5 \leq \Delta_1 \leq 0.5$ (which represents a variation of up to $\pm 50 \%$ on our initial guess with respect to the true value of $\tau(0)=1$), the results of our computations are very similar to the case of $\Delta_1=0$.

In the inset we investigate the effects of the initial conditions on the decentralized agreement protocol introduced in Sec.\ IV. We take the individual systems to be R\"ossler oscillators (\ref{ROSS}); we set $\Delta_1=0.8$ and plot the average estimation error $<E_2>$ versus the initial mismatch $\Delta_2$. As can be seen, for $-0.4 \lesssim \Delta_2 \lesssim 1.5$, the adaptive strategy is able  to produce good estimates of the unknown time-varying delay $\tau(t)$ within the time-span of our simulations ($2 \times 10^4$).
\begin{figure}[t]\label{IC}
\centering
\includegraphics{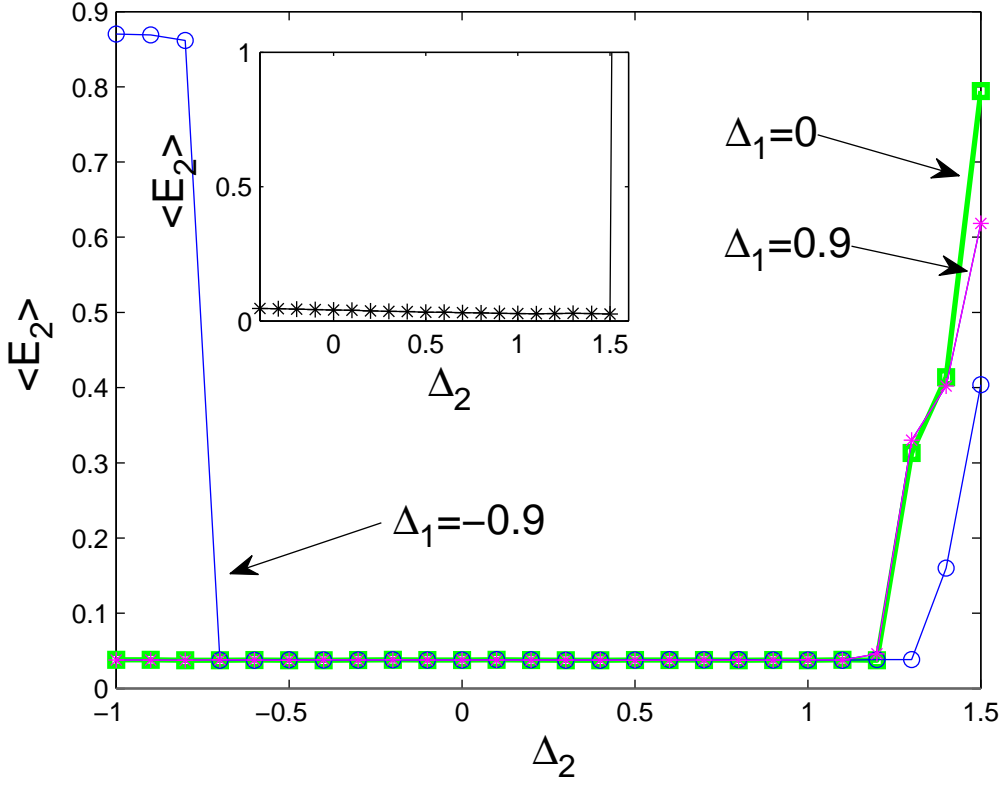}
\caption{\small We integrate the set of equations (\ref{main},\ref{adt},\ref{VDP}) with $\tau$ evolving according to Eq. (\ref{taut}) with $T=1$, $\zeta=0.5$, $\omega=10^{-3}$. The main plot shows the average estimation error $<E_2>$ versus the initial mismatch $\Delta_2$. Different markers are for different values of $\Delta_1$: circles correspond to $\Delta_1=-0.9$, squares correspond to $\Delta_1=0$, and  asterisks correspond to $\Delta_1=0.9$. Each point is an average over many different realizations (i.e., over different choices of the initial conditions for $x_1$ and $x_2$); $\alpha_1=\alpha_2=10^{-4}/2$,
$\gamma_1=\gamma_2=0.15$. The inset shows the results of simulations that study the effects of initial conditions on the decentralized agreement protocol introduced in Sec.\ IV. We integrate the set of equations (\ref{main},\ref{adtH},\ref{ROSS}) with $\tau$ evolving according to (\ref{taut}), $T=0.4$, $\zeta=0.2$, $\omega=10^{-3}$. We set $\alpha_1=\alpha_2=10^{-4}/2$,
$\gamma_1=0.6$, $\gamma_2=0.5$, $\kappa=1$. The inset shows the average estimation error $<E_2>$ versus the initial mismatch $\Delta_2$ for $\Delta_1=0.8$. }
\end{figure}

\section{Effects of noise}

In this section we consider the presence of noise in the communication channels between the two platforms. We assume additive noise and 
we replace $r_i$ in (\ref{main}b) by,
\begin{equation}
r_i(t)=  H(x_j(t-\tau(t)))+\sqrt{s} \sigma_i \epsilon_i(t), \label{rep}
\end{equation}
 where $\epsilon_i(t)$ is a zero-mean independent random number of unit variance drawn from a Gaussian distribution, $\sigma_i$ is a multiplicative factor, and $s$  is the time step of our integration method.


To better deal with the presence of noise in the received signals, we propose an alternative formulation of our adaptive strategy for which the potential (\ref{P}) is replaced by,
\begin{equation}
\tilde{\Psi}_i(t)=\nu \int^t e^{-\nu (t-\theta)} [r_i(\theta)-H(x_i(\theta-\tau'_i(\theta)))]^2 d\theta. \label{Psiti}
\end{equation}
From (\ref{Psiti}) we see that $\tilde{\Psi}_i$ is an exponential moving average of the squared synchronization error with averaging time $\nu^{-1}$. We require $\nu^{-1}$ to be larger than $T_x$, the characteristic timescale on which an uncoupled system evolves, and to be smaller than $T_p$, the timescale on which the communication-delay changes,
\begin{equation}\label{TS3}
T_x < \nu^{-1} < T_p.
\end{equation}
Note that $\tilde{\Psi}_i\geq 0$ and $\tilde{\Psi}_i=0$ only if $x_1=x_2$ and $\tau'_i=\tau$ (see Sec.\ IIA for a more detailed discussion).

Following our previous derivations, the equations for the alternative adaptive strategy are,
\begin{equation}
\dot{\tau'}_i=-\alpha_i \frac{\partial \tilde{\Psi}_i}{\partial \tau'_i}=-2 \alpha_i \xi_i, \label{tauti}
\end{equation}
where $\alpha_i>0$, and
\begin{equation}
\xi_i(t)= \nu \int^t e^{-\nu (t-\theta)} [r_i(\theta)-H(x_i(\theta-\tau'_i(\theta)))] DH \dot{x}_i(\theta-\tau'_i(\theta)) d\theta. \label{extra}
\end{equation}
We observe that the quantity $\xi_i$ in (\ref{extra}) obeys the following differential equation,
\begin{equation}
\dot{\xi}_i(t)=-\nu{\xi}_i(t)+ \nu [r_i(t)-H(x_i(t-\tau'_i(t)))] DH \dot{x}_i(t-\tau'_i(t)). \label{extrati}
\end{equation}
Thus our alternative adaptive strategy is completely described by the set of Eqs. (\ref{main}a,\ref{rep},\ref{tauti},\ref{extrati}). We note that with respect to the original formulation, the alternative strategy requires integration of an additional differential equation at each platform.


\begin{figure}[h!]\label{N2}
\centering
\includegraphics{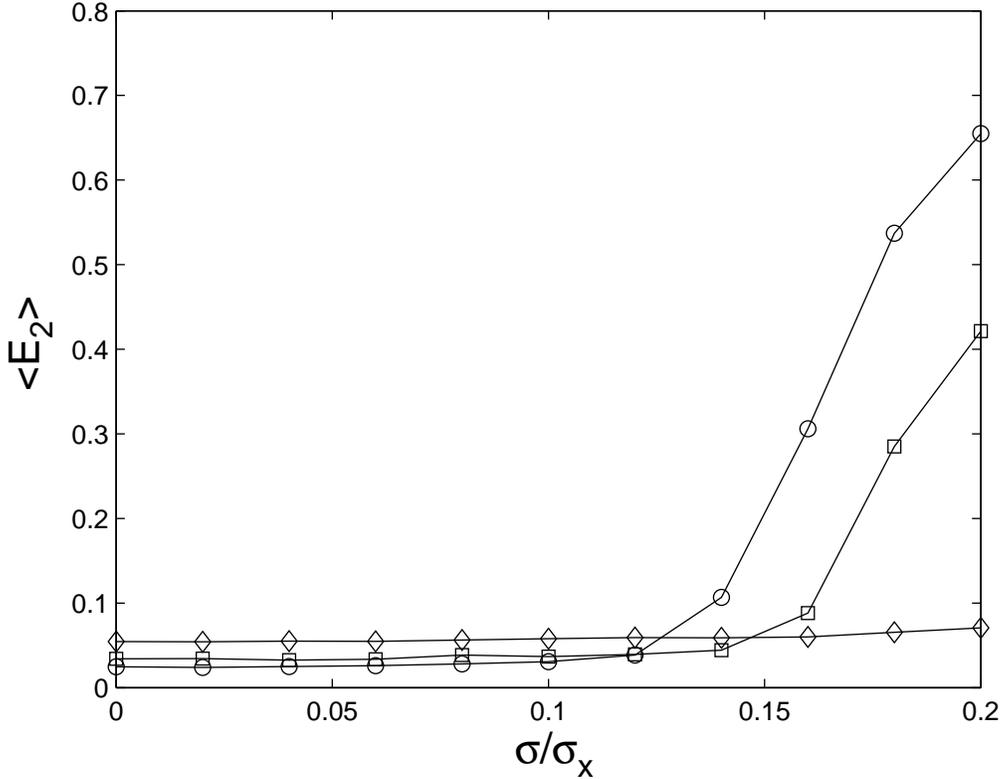}
\caption{\small The figure shows the results of numerical integration from $t=0$ to $t=10^4$ of the set of equations (\ref{main},\ref{ROSS},\ref{rep},\ref{adtHnu}),  corresponding to our alternative formulation of the adaptive strategy for autonomous systems, $H(x_i(t))=x_{i1}(t)$ and $\Gamma_i=[\gamma_i,0,0]$. $\tau(t)$ evolves according to Eq.\ (\ref{taut}), with $T=0.3$, $\zeta=0.15$, $\omega=10^{-3}$. We set  $\alpha_1=\alpha_2=10^{-4}/2$,
$\gamma_1=0.6$, $\gamma_2=0.5$, $\kappa=1$. We record the average estimation error $<E_2>$ for increasing values of the noise ratio $\sigma/\sigma_x$, where $\sigma=\sigma_1=\sigma_2$ and $\sigma_x=5.5$ is the standard deviation of the time evolution of $H(x_{s1})$.   Squares correspond to $\nu=1.5$ and diamonds to $\nu=3$. Circles correspond to runs in which the same simulation is repeated for our original formulation of the adaptive strategy, Eqs.\ (\ref{main},\ref{adt}).  Each point is an average over many different realizations, i.e., over different choices of the initial conditions for $x_1$ and $x_2$ (constant over the time-interval $[-\bar{\tau}_{max},0]$).}
\end{figure}


We have also considered the effects of noise on the decentralized agreement protocol introduced in Sec.\ IV. For this case,  Eq. (\ref{adtH}) is replaced by
\begin{equation}
\begin{split} \label{adtHnu}
\dot{\tau'}_i & = -2 \alpha_i \xi_i,  \\
\dot{\xi}_i & =  -\nu \xi_i +  \{ [r_i(t)-H(x_i(t-\tau'_i(t)))] DH(x_i(t-\tau'_i(t))) \dot{x}_i(t-\tau'_i(t)) -\kappa [\tilde{\tau}_j(t-\tau(t))-\tau'_i(t-\tau'_i(t))]  \},
\end{split}
\end{equation}
$i=1,2$, $j=3-i$, $\alpha_i>0$, $\tilde{\tau}_j(t-\tau(t))= {\tau'}_j(t-\tau(t))+\sqrt{s} \sigma_i \epsilon_i(t)$, with $s$, $\sigma_i$, and $\epsilon_i$ being the same as in Eq.\ (\ref{rep}).

In Fig.\ 7 we show the results of numerical experiments involving two coupled chaotic R\"ossler oscillators. We integrate the set of equations (\ref{main},\ref{ROSS},\ref{rep},\ref{adtHnu}),  corresponding to our alternative formulation of the adaptive strategy for autonomous systems. We consider that the communication-delay $\tau$ slowly varies in time according to Eq. (\ref{taut}),  with $T=0.3$, $\zeta=0.15$, $\omega=10^{-3}$ (see the figure caption for the simulation parameters).  We record the average estimation error $<E_2>$ for increasing values of the noise ratio $\sigma/\sigma_x$, where $\sigma=\sigma_1=\sigma_2$ and $\sigma_x=5.5$ is the standard deviation of the time evolution of $H(x_{s1})$.   Squares correspond to $\nu=1.5$ and diamonds to $\nu=3$. Circles correspond to runs in which the same simulation is repeated for our original formulation of the adaptive strategy, Eqs.\ (\ref{main},\ref{adtH}).
As can be seen, for low noise (low values of $\sigma$) the original adaptive strategy and the modified adaptive strategy (for both cases of  $\nu=1.5$ and $\nu=3$) produce similar results. As $\sigma$ is increased, the average estimation error $<E_2>$ grows. For large enough noise $\sigma$ ($\sigma \gtrsim 0.12 \sigma_x$), the modified adaptive strategy outperforms the original adaptive strategy.


\section{A strategy for simultaneously estimating the communication-delay and the signal amplitude}

In the previous sections we have shown that a suitable adaptive strategy based on synchronization of chaos can be used to estimate an unknown communication-delay between two coupled systems and achieve synchronization between them. As a reference application, we have proposed that such a strategy could be used to estimate the relative distance between two autonomously moving platforms.

Another aspect that we have not taken into account is that since communication occurs between two moving platforms, the amplitude of the signal received at each platform could be unknown and time-varying as well.  One viable strategy could be to try to infer the unknown amplitude from the delay. In this section, we will show that the adaptive strategy proposed in Secs. IV can be effectively extended to simultaneously and independently estimate the unknown communication-delay and the unknown amplitude, even when they evolve independent of each other and in a noisy environment. We wish to emphasize that an adaptive strategy to estimate the amplitude of the signal alone has already been studied in \cite{SOTT,SOTT2} and experimentally tested in \cite{EXP,MurphyEtAl,EXP2}.

In order to take into account the effect of the unknown time-varying amplitude of the signal received at each platform, we replace equation (\ref{main}) by the following,
\begin{subequations}\label{ID2}
\begin{align}
\dot{x}_i(t)= & F(x_i(t))+\Gamma_i[r_i(t)- A'_i(t) H(x_i(t-\tau'_i(t)))], \\
r_i(t)=  & A(t) H(x_j(t-\tau(t))), 
\end{align}
\end{subequations}
$i=\{1,2\}$, $j=(3-i)$, where both the amplitude $A(t)$ and the communication-delay $\tau(t)$ are unknown and slowly-evolving in time and $A'_i(t)$ is an estimate at platform $i$ of the unknown amplitude $A(t)$. Again, we assume that the two platforms are coupled in line of sight, so that it can be assumed that the attenuation of the signal from platform $i$ to platform $j$ is the same as that from platform $j$ to platform $i$ and that the communication-delay is the same in both directions.

We now seek to obtain an adaptive strategy to simultaneously and independently estimate at each platform both the unknown amplitude $A(t)$ and the communication delay $\tau(t)$. We assume that the two chaotic systems at each platform are autonomous, thus we use a decentralized agreement protocol similar to the one presented in Sec. IV. Similarly to Sec. IV, we introduce at each platform $i$ a potential,
\begin{equation}
\hat{\Psi}_i(t)=[A(t) H(x_j(t-\tau(t)))- A'_i(t) H(x_i(t-\tau'_i(t)))]^2+ \kappa [A(t) \tau'_j(t-\tau(t))- A'_i(t) \tau'_i(t-\tau'_i(t))]^2. \label{Psibis}
\end{equation}
Note that $\hat{\Psi}_i\geq 0$ and $\hat{\Psi}_i=0$ only if $x_1=x_2$, $\tau'_i=\tau$, and $A'_i=A$. The latter follows from the observation that $H(x_i)$ and $H(x_j)$ are chaotic and rapidly evolving in time, while $A(t)$ and $A'_i(t)$ can be considered constant over the timescale on which chaos evolves. Then the only way that the first term on the right hand-side of (\ref{Psibis}) can be set  equal to zero independent of time is for $A'_i=A$, for which case $A'_i=A$ can be factored out of the square brackets in (\ref{Psibis}). Then the condition for $\hat{\Psi}_i=0$ is that both the terms in the square brackets are equal zero, yielding $x_1=x_2$, and $\tau'_i=\tau$ (for more details, see Sec. IIA).

Thus we seek to formulate an adaptive strategy that seeks to evolve $A'_i(t)$ and $\tau'_i(t)$ in such a way to minimize the potential (\ref{Psibis}). To this aim, we introduce the following gradient descent relations,
\begin{subequations}\label{GDAT}
\begin{align}
\dot{A'}_i=  -\beta_i \frac{\partial \hat{\Psi}_i}{\partial A'_i}= 2 \beta_i  \{& [A(t)  H(x_j(t-\tau(t)))- A'_i(t) H(x_i(t-\tau'_i(t)))]H(x_i(t-\tau'_i(t))) \nonumber \\ & +\kappa     [A(t) \tau'_j(t-\tau(t))- A'_i(t) \tau'_i(t-\tau'_i(t))]  \tau'_i(t-\tau'_i(t)) \}, \\
\dot{\tau'}_i=  -\alpha_i \frac{\partial \hat{\Psi}_i}{\partial \tau'_i}= -2 \alpha_i  \{& [A(t) H(x_j(t-\tau(t)))-A'_i(t) H(x_i(t-\tau'_i(t)))] A'_i(t) DH(x_i(t-\tau'_i(t))) \dot{x}_i(t-\tau'_i(t)) \nonumber \\ & -\kappa  [A(t) \tau'_j(t-\tau(t))- A'_i(t) \tau'_i(t-\tau'_i(t))] A'_i(t)\},
\end{align}
\end{subequations}
with $\alpha_i,\beta_i>0$, $i=\{1,2\}$.

We have numerically tested the adaptive strategy described by equations (\ref{ID2},\ref{GDAT}) with $\tau(t)$ evolving according to Eq.\ (\ref{taut}) with $T=0.3$, $\zeta=0.15$ and $A(t)$ evolving according to the following equation,
\begin{equation}
A(t)=A+ \eta \sin(\omega_A t), \label{aut}
\end{equation}
with $A=1$, $\eta=0.1$, $\omega_A=5 \times 10^{-4}$. We choose each individual system to be a chaotic R\"ossler  oscillator (\ref{ROSS}), $H(x_i(t))=x_{i1}(t)$ and $\Gamma_i=[\gamma_i,0,0]$. We set  $\alpha_1=\alpha_2=\beta_1=\beta_2=10^{-4}$,
$\gamma_1=0.6$, $\gamma_2=0.5$, $\kappa=1$. The results of our numerical experiment are shown in Fig.\ 8. As can be seen, after a transient the adaptive strategy is able to track the time-evolutions of both $\tau(t)$ and $A(t)$. Also, the time evolutions of $x_1(t)$ and $x_2(t)$ are synchronized (not shown). Note that the strategies to estimate the unknown delay and amplitude (given in Eqs. (\ref{GDAT}a) and (\ref{GDAT}b)) are run simultaneously and independently of each other.

\begin{figure}[h!]\label{ATT}
\centering
\includegraphics{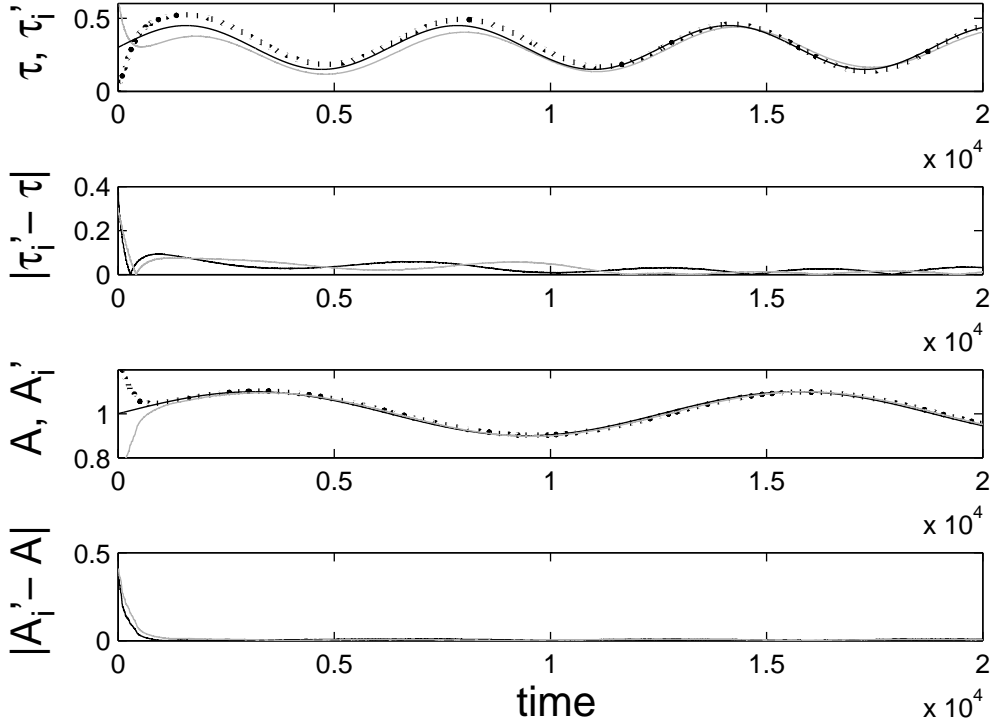}
\caption{\small The figure shows the results of numerical integration from $t=0$ to $t= 2 \times 10^4$ of the set of equations (\ref{ID2},\ref{ROSS},\ref{GDAT}),  corresponding to the version of the adaptive strategy for simultaneously estimating the communication-delay and the signal amplitude, $H(x_i(t))=x_{i1}(t)$ and $\Gamma_i=[\gamma_i,0,0]$. $\tau(t)$ evolves according to Eq.\ (\ref{taut}), with $T=0.3$, $\zeta=0.15$; $A(t)$ evolves according to Eq.\ (\ref{aut}), with $A=1$, $\eta=0.1$, $\omega_A=\omega/2=5 \times 10^{-4}$.  We set  $\alpha_1=\alpha_2=\beta_1=\beta_2=10^{-4}$,
$\gamma_1=0.6$, $\gamma_2=0.5$, $\kappa=1$. From top to bottom:  time evolution of $\tau(t)$ (thin black curve), $\tau'_1(t)$ (thin grey curve), and $\tau'_2(t)$ (thick dotted black curve); time evolution of the delay estimation errors, $|\tau(t)-\tau'_1(t)|,|\tau(t)-\tau'_2(t)|$; time evolution of $A(t)$ (thin black curve), $A'_1(t)$ (thin grey curve), and $A'_2(t)$ (thick dotted black curve); time evolution of the amplitude estimation errors, $|A(t)-A'_1(t)|,|A(t)-A'_2(t)|$. We select the initial conditions for $x_{i1},x_{i2},x_{i3}$ to be constant over the time-interval $[-\bar{\tau}_{max},0]$, where $\bar{\tau}_{max}=\max_t \tau_{max}(t)$, and equal to randomly chosen points on the R\"ossler chaotic attractor, $\tau'_1(0)=0.7$, $\tau'_2(0)=0$, $A'_1(0)=0.6$, $A'_2(0)=1.4$. }
\end{figure}

\section{Discussion}

Our strategies described in this paper take advantage of the properties of chaotic sequences of never repeating in time. Another way of generating long non-repeating time-traces is to consider sequences of independent random numbers.  For example, consider the $N$-random sequence
\begin{equation} \label{SN}
S(n)=\left\{\begin{array}{ll} 0,&  \mbox{with probability $p$,} \\ 1,& \mbox{with probability $1-p$.} \end{array} \right.
\end{equation}
$n=0,1,...,N$, with $0 \leq p\leq 1$. Then if we select two s-independent random sequences as defined in (\ref{SN}), we can compute the probability $P$ that they are the same, $P=(1 - 2p(1-p))^N$. It is easy to see that unless $p=0$ or $p=1$, $P$ tends to $0$ as $N$ tends to infinity. Hence it is very unlikely to produce two random sequences that are the same.

A way of getting around this problem, is to rely on a pseudorandom number generator, that is, a deterministic algorithm for generating a sequence of numbers that approximates the properties of random numbers. Such an algorithm requires initialization by a relatively small set of initial values, called the \emph{seed}. Two sequences that are initialized by the same seed are identical. The seed plays the same role as the initial condition of a chaotic dynamical system and indeed a way of producing pseudorandom sequences is to use an underlying deterministic chaotic process \cite{Uchida,Mu:Ro,Ka:Av:Re:Co:Ro}. In this respect, the use of a pseudorandom algorithm could represent a valid alternative to the use of a chaotic dynamical system for the delay-identification problem discussed in this paper.

However, the key-property that we present and exploit in this paper is that two chaotic dynamical systems, even if initialized from two different initial conditions, can be {synchronized}, i.e., by introducing an appropriate feedback mechanism, they can be maintained in a stable synchronized chaotic time-evolution \cite{Replace,He:Va,Kap}. Furthermore, it has been shown that even in the presence of slight non-identicality between the coupled systems and slight deviations from nominal conditions, a stable approximately synchronous evolution can be reached and maintained in time \cite{Ash1,Ash2,Gh:Ch,restr_bubbl,Su:Bo:Ni,So:Po}. Thus the use of a closed-loop architecture has the following advantages:

(i) The two systems do not need to be initialized with the same initial condition, i.e., $\emph{a-priori}$ knowledge of a \emph{seed} is not required.

(ii) The mechanism is robust with respect to noise and slight deviations from nominal conditions, such as slight non-identicality in the individual systems, slight variations in the amplitude of the received signal, or slightly different environmental conditions.

To conclude, the advantage of our chaos-synchronization strategy is that it relies on a closed-loop architecture that allows to maintain synchronization between two distant systems even in the presence of noise and non-identicality in the individual systems or mismatches in the initial conditions.

{On the other hand, noise or small mismatches in the parameters of the individual chaotic systems being coupled can be responsible for the onset of \emph{bubbling} \cite{bub1,bub2,restr_bubbl,SAS}, i.e., rare intermittent large deviations (\emph{bursts}) from synchronization. {Bubbling occurs when the synchronized state is stable for typical chaotic orbits but is unstable for certain unstable periodic orbits within the synchronized chaotic attractor. Consider for example the set of Eqs. (\ref{main},\ref{adtH}). For the case of constant $\tau$, these equation allow an invariant set, the so-called synchronous manifold (SM), defined as $x_1(t)=x_2(t)$, $\tau'_1=\tau'_2=\tau$. Stability of the SM with respect to infinitesimal transversal perturbations can be quantified in terms of the maximum transverse Lyapunov exponent (MTLE) of the system.  Now assume the MTLE is negative, implying that stability is observed for any initial condition on the attractor. However, in the presence of either noise, or small mismatches in the parameters of the individual systems, or even a slowly time-varying $\tau(t)$, it is possible that the trajectory eventually gets close to an unstable periodic orbit embedded in the attractor having an associated positive MTLE. If this happens, the trajectory may be repelled away from the SM  and  eventually return close to the SM after some time, giving rise to a burst. This phenomenon is called \emph{bubbling} \cite{bub1,bub2}}.
 In \cite{SAS} bubbling is observed for a network of coupled dynamical systems, each of which independently implements an adaptive strategy to maintain synchronization.
Analogously, we expect bubbling to eventually arise (depending on the choice of the equation parameters), for the problem described by Eqs. (\ref{main},\ref{adtH}), i.e., for a case in which two coupled systems implement an adaptive strategy to estimate a communication-delay. In particular, bubbling is likely to arise in the case in which the unknown delay is time-varying, as mismatches in the delay estimate $|{\tau'}_i(t)-\tau(t)|$, no matter how small they are, may generate it.
}

\section{Conclusions}

In this paper, we considered synchronization of two bidirectionally coupled chaotic systems that communicate with an unknown time-varying delay. We presented a nonlinear adaptive strategy to simultaneously identify the time-varying communication delay and achieve synchronization.

To achieve the twofold goal of synchronizing the chaotic systems and identifying the unknown delay, we proposed an adaptive strategy based on the minimization of an appropriately defined potential through a gradient descent technique (see also \cite{SOTT,SOTT2,EXP}). 
The proposed strategy is presented in two alternative formulations to deal with either non-autonomous or autonomous systems. The version for non-autonomous systems requires the presence of two perfectly synchronized internal clocks at the two platforms, while the version for autonomous systems is based on a decentralized agreement protocol, i.e., on exchanging information on the respective delay-estimates  between the two platforms.

As a reference application, we have considered the problem of identifying the communication-delays between two autonomous moving platforms. We have assumed that two identical chaotic oscillators are installed at each platform and that these chaotic systems seek to synchronize via a signal broadcast from one platform to the other (and viceversa). 
Moreover, we have assumed that the signal is transmitted with an unknown communication-delay that depends on the distance between the platforms. Our strategy could be used to dynamically estimate at each platform the distance at which the other platform is at any given time.

The effectiveness of the proposed approach has been illustrated by means of extensive numerical simulations, showing that the adaptive strategy can be effective in synchronizing the chaotic systems and in correctly estimating the time-varying communication-delay.  Another advantage of our approach is that in addition to estimating the communication-delay, the adaptive strategy could be used to simultaneously identify other parameters (see e.g., \cite{SOTT,SOTT2,IDTOUT,IDDD}). {For example, in Sec. VII we have shown that the adaptive strategy is effective in simultaneously and independently estimating the unknown time-varying communication delay and the unknown time-varying amplitude of the received signal.}

To investigate the robustness of the proposed approach, we have numerically tested the performance of our strategy with respect to the choice of the initial conditions. We have also investigated the effects of noise in the communication channels between the platforms and proposed an appropriately modified adaptive strategy for which a relevant enhancement of the performance has been observed in the presence of noise.

{Our adaptive strategy (especially in its version for nonautonomous systems) could be useful in applications as different as the identification of communication delays between moving platforms, clock synchronization, the estimation of propagation delays in an asynchronous code-division multiple access communication system, and the localization of a mobile target through a bistatic radar system. However, a main limitation to our approach seems to be that for some applications (e.g., clock synchronization in a computer network) the communication delays may be asymmetrical (i.e., different in the two directions from $i$ to $j$ and from $j$ to $i$). The condition of symmetrical delays, which holds for  line of sight communication, seems to be relevant to the formulation of our strategy that applies to autonomous systems. How to extend the strategy for autonomous systems to the case of asymmetrical delays is the subject of ongoing investigations. One viable approach would be to increase the amount/type of information that the platforms are allowed to exchange. As a reference for future work, we also note the importance of testing our strategy with systems displaying larger complexity than the simple Van der Pol and R\"ossler oscillators considered in this paper (e.g., systems of interest for practical applications).}

{\centering{ACKNOWLEDGEMENTS}}

Francesco Sorrentino thanks Ed Ott, Nick Mecholsky, Bahargava Ravoori, Adam Cohen, Thomas Murphy, and Raj Roy for insightful discussions.
Both authors are thankful to Mario di Bernardo for his help and support. Both authors are indebted to the reviewers of the paper for their sensible comments.

\end{document}